\documentclass[12pt,a4paper]{article}
\usepackage{setspace}
\usepackage{subfig}
\usepackage{authblk}

\usepackage{listings}
\usepackage{caption}
\usepackage{amsmath, epsfig,graphicx}
\usepackage{color}
\usepackage{enumerate}

\def\Title#1{\begin{center} {\Large \bf #1 } \end{center}}
\def\Abstracti#1{\begin{center} {\small \bf #1 } \end{center}}
\def\Author#1{\begin{center}{ \sc #1} \end{center}}
\def\Address#1{\begin{center}{ \it #1} \end{center}}
\newenvironment{Abstract}{\begin{quotation}  }{\end{quotation}}

\begin{document}

\begin{titlepage}
\let\thefootnote\relax\footnote{* Mansouri@quantgates.co.uk}

\Title{}
\Title{}
\Title{An Analysis of the Quantum Approximation Optimisation Algorithm}
\vfil
\Author{ Behzad Mansouri*}
\vfil
\Address{QuantGates Ltd, 
London, EC2A 4NE, UK}
\vfil
\Abstracti {Abstract}
\begin{Abstract}
This article consists of a short introduction to the quantum approximation optimisation algorithm (QAOA).
The mathematical structure of the QAOA, as well as its basic properties, are described. The implementation of the QAOA on MaxCut 
problems, quadratic unconstrained binary optimisation problems (QUBOs), and Ising-type Hamiltonians is considered in detail.

\end{Abstract}
\vfill

\end{titlepage}

\section{\large The Quantum Approximate Optimisation Algorithm}
The quantum approximation optimisation algorithm (QAOA) was first constructed and implemented on MaxCut 
problems by Farhi et al. \cite{1}. The QAOA was specifically designed to run on circuit-based quantum computers. The quality of the approximation provided by the algorithm directly depends on an integer $p$, and with $p$ growing slowly with $n$ (the input size, $n$ bits) we hope the algorithm will be powerful enough to find solutions beyond the capabilities of classical algorithms. In \cite{2}, the QAOA was applied to E3LIN2 (MAX-3XOR) and initially beat the state-of-the-art classical algorithm. Shortly after, a better classical algorithm was constructed that outperformed the QAOA in this case \cite{3}. Besides these empirical results, \cite{4} argued that the QAOA can exhibit a form of quantum supremacy in such a way that even the lowest depth of the algorithm cannot be simulated efficiently using any classical computer, based on, of course, reasonable assumptions in complexity theory.
Regarding the implementation of the QAOA for a variety of combinatorial optimisation problems, Hadfield et al. \cite{5} provided a framework for designing  QAOA circuits for these problems, including graph colouring, the travelling salesman problem (TSP), and single machine scheduling (SMS). Moreover, the QAOA has been extended by introducing the quantum alternating operator ansatz \cite{6} to encompass a more general class of
quantum states by allowing alternation between broader classes of operators in which it can widen the applicability of the QAOA.
Herein, we reviewed how the QAOA works and its implementation on MaxCut problems, and finally, we studied the implementation of this promising algorithm on  quadratic unconstrained binary optimisation problems (QUBOs) and Ising-type Hamiltonians, in which the QUBOs embrace many combinatorial optimisation problems, showing that it has practical applications in every industry.

The QAOA is mainly designed to tackle constraint satisfaction problems (CSPs). A large number of combinatorial optimisation problems can be recast in the form of constrained satisfaction problems, 
the latter of which can be defined with $n$ bits and $m$ constraints.

The objective function of a CSP can be written as:
\begin{equation}
C(z) = \sum\limits_{a = 1}^m {{C_a}} (z)
\end{equation}

where $z = {z_1}{z_2}...{z_n}$ is the bit string and ${C_a}(z) = 1$ if $z$ satisfies the constraint $a$, and 0 otherwise \cite{1}.
By promoting $n$ bits to $n$ qubits, we can construct a ${2^n}$ dimensional Hilbert space with a computational basis $\left| z \right\rangle$. The constraint functions in the above can be written in diagonal matrix forms. By promoting $(1)$ as an operator (diagonal form) acting on Hilbert space vectors, we have:

\begin{equation}
C\left| z \right\rangle  = \sum\limits_{a = 1}^m {{C_a}} \left| z \right\rangle  = \sum\limits_{a = 1}^m {{P_a}} \left| z \right\rangle  = P(z)\left| z \right\rangle 
\end{equation}

where ${{P_a}}$ is an eigenvalue corresponding to the constraint operator ${{C_a}}$. Here, the largest value in the eigenvalue set, $P(z')$, is the maximum of the objective function $C$. When considering a general state in our Hilbert space as $\left| \psi  \right\rangle $, one can evaluate the expectation value of the operator $C$ as below:

\begin{equation}
\begin{array}{*{20}{l}}
{\left\langle C \right\rangle  = \langle \psi |C\left| \psi  \right\rangle  = (\sum\limits_{z}^{} {{b_z}^*} \langle z|)(\sum\limits_{z}^{} {{b_z}} C\left| z \right\rangle )}\\
{ = \sum\limits_{z}^{} {{{\left| {{b_z}} \right|}^2}P(z) \le } \sum\limits_{z}^{} {{{\left| {{b_z}} \right|}^2}P(z')}  = P(z')}
\end{array}
\end{equation}

The maximum of $\left\langle C \right\rangle $ is equal to $P(z')$, the eigenvalue associated with the eigenvector $\left| {z'} \right\rangle $. This setup allows us to create our algorithm. The operator $C$ with its eigenvalues will $mark$ the computational basis into the two search problem categories of $solutions$ and $not$-$solutions$, 
and the problem of finding the solutions will be addressed by maximising of the expectation value of the objective function $C$ in a $p$-layered variational ansatz. To achieve this aim, we review Farhi's prescription below.

First, a unitary operator is constructed as:
\begin{equation}
U(C,\gamma ) = {e^{ - i\gamma C}} = \prod\limits_{a = 1}^m {{e^{ - i\gamma {C_a}}}}  
\end{equation}

As $[{C_a },{C_b }] = 0$, it can be written as the product of exponentials.

This quantum gate can be applied on a general state $\left| \psi  \right\rangle $; then, we get:

\begin{equation}
\begin{array}{l}
U(C,\gamma )\left| \psi  \right\rangle  = \sum\limits_{z}^{} {{e^{ - i\gamma \sum\limits_{a = 1}^m {{C_a}} }}{a_z}} \left| z \right\rangle \\= \sum\limits_{z }^{} {(\prod\limits_{a = 1}^m {U({C_a }} ,\gamma ){a_z}} \left| z \right\rangle 
\end{array}
\end{equation}

The diagonal constraint operators allow to write $U(C,\gamma)$ as the product of exponentials, and this makes the implementation of the gate via universal gates simpler.
Note that here, for the eigenbasis $\left| {z'} \right\rangle $ satisfied by constraints ${C_a }$, a phase change (${e^{ - i\gamma }}$) is received, and this can make the ${\left| {\left. {a'} \right|} \right.^2}$ large enough, and obviously, by the corresponding measurement, we can obtain the state with the higher probability. One can implement such quantum gates with the help of controlled phase gates and one ancilla qubit.

As an explicit example, consider $\left| \psi  \right\rangle  = \left| {{y_1}} \right\rangle \left| {{y_2}} \right\rangle $, where $\left| {{y_1}} \right\rangle  = {\alpha _0}\left| 0 \right\rangle  + {\alpha _1}\left| 1 \right\rangle $ and $\left| {{y_2}} \right\rangle  = {\beta _0}\left| 0 \right\rangle  + {\beta _1}\left| 1 \right\rangle $, and two constraints as ${{C_1} = 1}$ if ${\left| {{y_1}} \right\rangle  = \left| {{y_2}} \right\rangle  = \left| 1 \right\rangle }$ and ${{C_2} = 1}$ if $\left| {{y_1}} \right\rangle  = \left| 1 \right\rangle $

First, we directly applied $U(C,\gamma)$ to our two qubit systems.
\begin{equation}
\begin{array}{l}
U(C,\gamma )\left| {{y_1}{y_2}} \right\rangle  = {e^{ - i\gamma {C_2}}}{e^{ - i\gamma {C_1}}}({\alpha _0}{\beta _0}\left| {00} \right\rangle  + {\alpha _0}{\beta _1}\left| {01} \right\rangle  + {\alpha _1}{\beta _0}\left| {10} \right\rangle  + {\alpha _1}{\beta _1}\left| {11} \right\rangle )\\
= ({\alpha _0}{\beta _0}\left| {00} \right\rangle  + {\alpha _0}{\beta _1}\left| {01} \right\rangle  + {e^{ - i\gamma }}{\alpha _1}{\beta _0}\left| {10} \right\rangle  + {e^{ - 2i\gamma }}{\alpha _1}{\beta _1}\left| {11} \right\rangle )
\end{array}
\end{equation}

Now, with help of two controlled phase gates and one ancilla qubit, we obtained an equivalent result.
\begin{figure}
		\centering
	\includegraphics[width=3in]{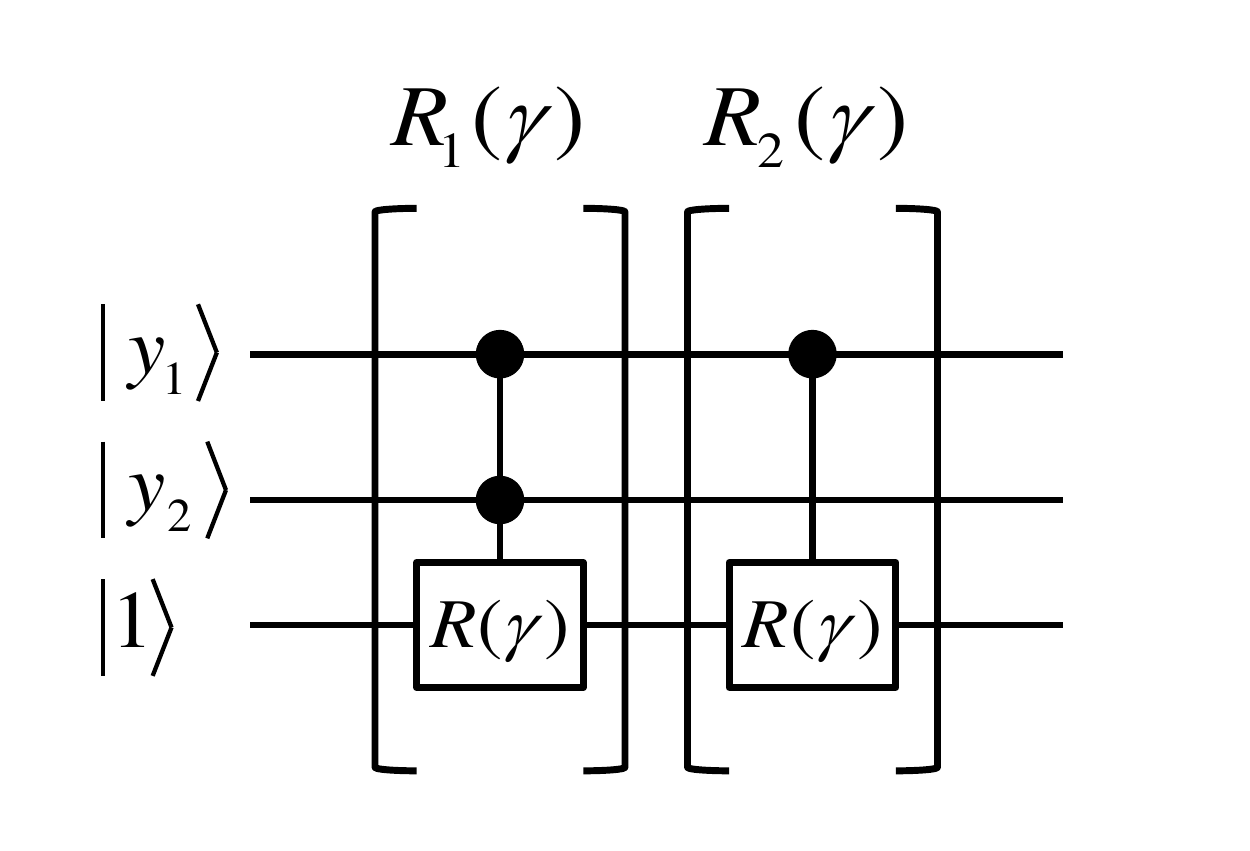}
	\caption{Implementation using two controlled-phase gates and  one ancilla qubit.}

\end{figure}

Here, we define our phase gate:
$R(\gamma ) = \left[ {\begin{array}{*{20}{c}}
	1&0\\
	0&{{e^{-i\gamma }}}
	\end{array}} \right]$

The ancilla qubit was prepared in $\left| 1 \right\rangle $, resulting in:
\begin{equation}
\left| {{y_1}{y_2}} \right\rangle \left| 1 \right\rangle  = {\alpha _0}{\beta _0}\left| {00} \right\rangle \left| 1 \right\rangle  + {\alpha _0}{\beta _1}\left| {01} \right\rangle \left| 1 \right\rangle  + {\alpha _1}{\beta _0}\left| {10} \right\rangle \left| 1 \right\rangle  + {\alpha _1}{\beta _1}\left| {11} \right\rangle \left| 1 \right\rangle 
\end{equation}

Consider ${R_1}(\gamma)$ and ${R_2}(\gamma)$ in their explicit matrix form:

\begin{equation}
{R_1}(\gamma ) = \left[ {\begin{array}{*{20}{c}}
	I&0&0&0\\
	0&I&0&0\\
	0&0&I&0\\
	0&0&0&{R(\gamma )}
	\end{array}} \right]\begin{array}{*{20}{c}}
{}&{}
\end{array}{R_2}(\gamma ) = \left[ {\begin{array}{*{20}{c}}
	I&0&0&0\\
	0&I&0&0\\
	0&0&{R(\gamma )}&0\\
	0&0&0&{R(\gamma )}
	\end{array}} \right]
\end{equation}

Here, the ${R_1}(\gamma)$ gate controls two qubits, $\left| {{y_1}} \right\rangle $ and $\left| {{y_2}} \right\rangle $, and acts on our third qubit (auxiliary qubit) as its target. Based on our constraints here, the gate ${R_1}(\gamma)$ only acts $R(\gamma)$ on ancilla when we have $\left| {{y_1}} \right\rangle  = \left| {{y_2}} \right\rangle  = \left| 1 \right\rangle $ and we get
\begin{equation}
{R_1}(\gamma )\left| {{y_1}{y_2}} \right\rangle \left| 1 \right\rangle  = {\alpha _0}{\beta _0}\left| {00} \right\rangle \left| 1 \right\rangle  + {\alpha _0}{\beta _1}\left| {01} \right\rangle \left| 1 \right\rangle  + {\alpha _1}{\beta _0}\left| {10} \right\rangle \left| 1 \right\rangle  + {\alpha _1}{\beta _1}\left| {11} \right\rangle  \otimes {R}(\gamma )\left| 1 \right\rangle 
\end{equation}

After that, ${R_2}(\gamma)$ gate comes into play in which it controls only one qubit $\left| {{y_1}} \right\rangle $ and acts on the third qubit as its target, resulting in:

\begin{equation}
\begin{array}{l}
{R_2}(\gamma ){R_1}(\gamma )\left| {{y_1}{y_2}} \right\rangle \left| 1 \right\rangle \\
= {\alpha _0}{\beta _0}\left| {00} \right\rangle \left| 1 \right\rangle  + {\alpha _0}{\beta _1}\left| {01} \right\rangle \left| 1 \right\rangle  + {\alpha _1}{\beta _0}\left| {10} \right\rangle  \otimes {R}(\gamma )\left| 1 \right\rangle  + {\alpha _1}{\beta _1}\left| {11} \right\rangle  \otimes {R}(\gamma ){R}(\gamma )\left| 1 \right\rangle \\
= ({\alpha _0}{\beta _0}\left| {00} \right\rangle  + {\alpha _0}{\beta _1}\left| {01} \right\rangle  + {e^{ - i\gamma }}{\alpha _1}{\beta _0}\left| {10} \right\rangle  + {e^{ - 2i\gamma }}{\alpha _1}{\beta _1}\left| {11} \right\rangle )\left| 1 \right\rangle 
\end{array}
\end{equation}

Note that here, based on the simple nature of our explicit constraints, the circuit implementation is straightforward. In general, more complicated constraints require us to apply more complicated gates.

Note ${e^{ - ik\gamma }}{a_z}\left| z \right\rangle $ is equal to ${a_z}\left| z \right\rangle $ up to the phase factor ${e^{ - ik\gamma }}$, and the output of the measurement for these two states are the same. In fact, the unitary operator $U(C,\gamma)$ is diagonal in the computational basis and does not push the probability around. The origin of this diagonality comes from selecting a specific type of optimisation problem (i.e., CSP) in which the computational basis is taken to be the eigenbasis of the constraint operators. In order to save and see the effect of this phase change related to $U(C,\gamma)$ in our measurement, we needed to introduce a rotation operator that takes $\left| 0 \right\rangle$ to a combination of $\left| 0 \right\rangle$ and $\left| 1 \right\rangle$; inevitably, this operator will increase the degrees of freedom by one at the lowest depth. A convenient choice of such operator can be given by:
\begin{equation}
U(B,\beta ) = {e^{ - i\beta B}} = \prod\limits_{i = 1}^n {{e^{ - i\beta {\sigma _i}^x}}} 
\end{equation}
With operator $B$ defined as
\begin{equation}
B = \sum\limits_{i = 1}^n {{\sigma _i}^x}  = {\sigma _1}^x \otimes {I^{ \otimes (n - 1)}} + {I_1} \otimes {\sigma _2}^x \otimes {I^{ \otimes (n - 2)}} + ... + {I^{ \otimes (n - 1)}} \otimes {\sigma _n}^x
\end{equation}

\begin{figure}
	\centering
	\includegraphics[width=2.5 in]{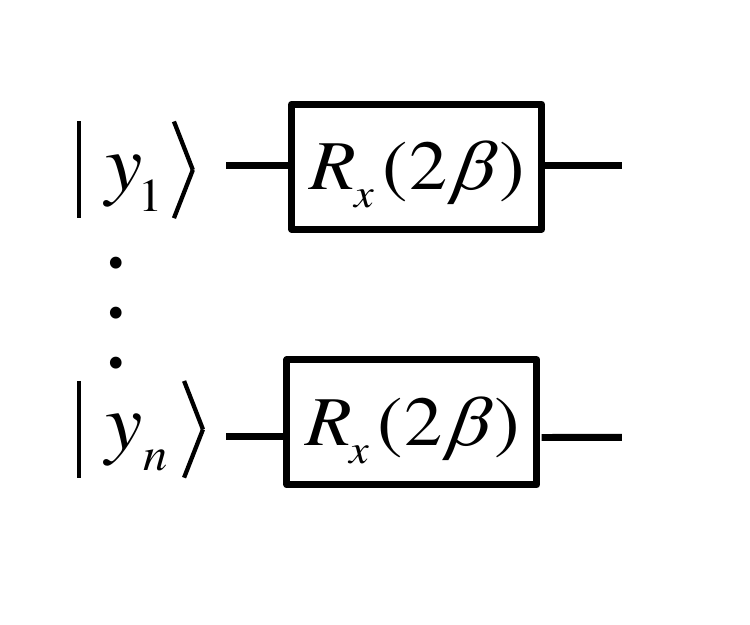}
	\caption{$U(B,\beta )$ implementation with depth 1}
	
\end{figure}

Again, as $[{\sigma _i}^x \otimes {I_j},{I_i} \otimes {\sigma _j}^x] = 0$, $U(B,\beta)$ can be written as the product of exponentials.

Let us see the effect of $U(B,\beta)$ on an explicit example of one qubit.
Consider a qubit in its superposition, $\left| y \right\rangle  = a\left| 0 \right\rangle  + b\left| 1 \right\rangle $, and suppose $U(C,\gamma)$ delivers a phase factor ${e^{ - i\gamma }}$ to base $\left| 0 \right\rangle $. Now, applying ${e^{ - i\beta {\sigma ^x}}}$ and performing measurements, we can obtain a probability of ${a^2}{\cos ^2}(\beta) + {b^2}{\sin ^2}(\beta) + ab\sin (2\beta)\sin (\gamma)$ to get $\left| 0 \right\rangle $, and we can see the contribution of the phase, related to $U(C,\gamma)$, as desired.

By acting the operator $U(B,{\beta _\alpha })U(C,{\gamma _\alpha })$ $p$ times with different ${\gamma _\alpha }$ and ${\beta _\alpha }$, we arrive at the below  $p$-layered variational ansatz:

\begin{equation}
\left| {\vec \gamma ,\vec \beta } \right\rangle  = U(B,{\beta _p})U(C,{\gamma _p})...U(B,{\beta _1})U(C,{\gamma _1})\left| s \right\rangle 
\end{equation}

Where $\left| s \right\rangle  = \frac{1}{{{2^{n/2}}}}\sum\limits_z {\left| z \right\rangle }$ is the uniform superposition state.
Now consider the expectation of C in this quantum state
\begin{equation}
F(\vec \gamma ,\vec \beta ) = \left\langle {\vec \gamma ,\vec \beta } \right|C\left| {\vec \gamma ,\vec \beta } \right\rangle 
\end{equation}

\begin{figure}
	\centering
	\includegraphics[width=4.5in]{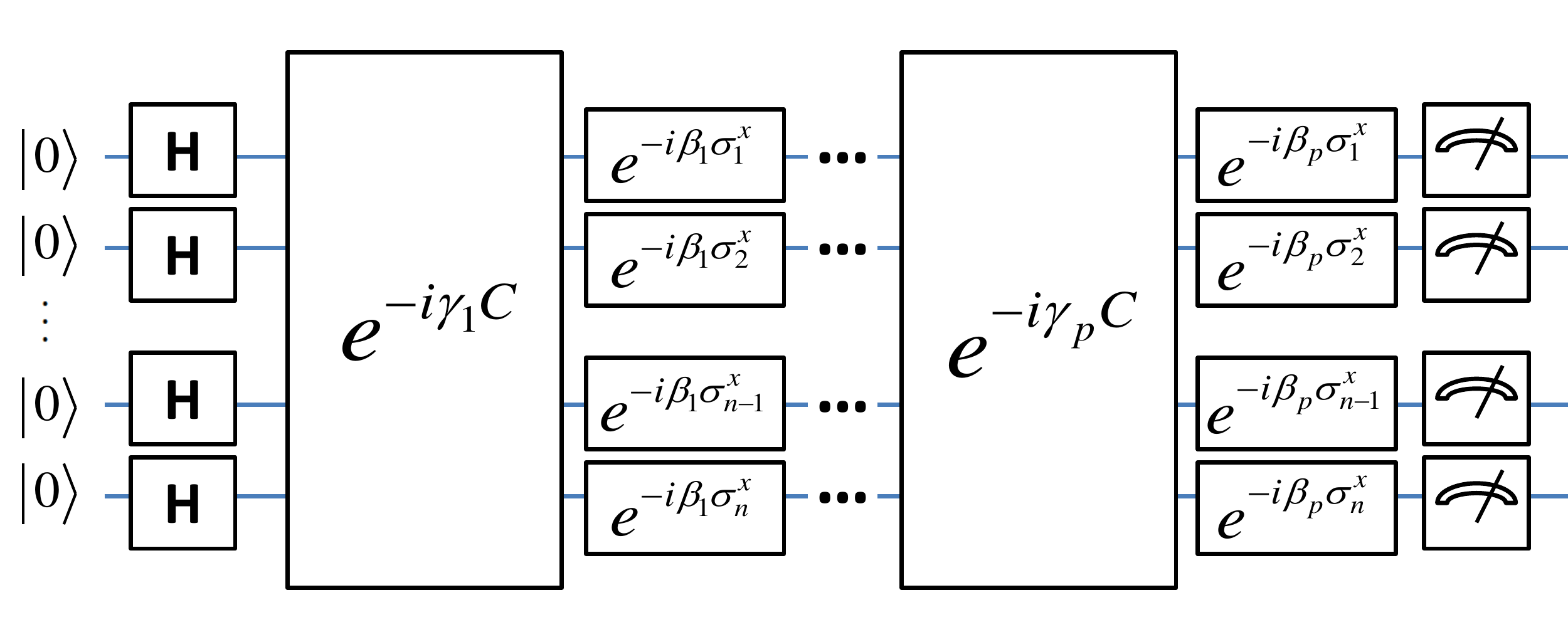}
	\caption{The QAOA circuit diagram.}
	\label{fig:boat1}
\end{figure}

\begin{equation}
\begin{array}{l}
	\left\langle {\vec \gamma ,\vec \beta } \right|C\left| {\vec \gamma ,\vec \beta } \right\rangle  = \left\langle s \right|{U^\dag }(C,{\gamma _1})...{U^\dag }(B,{\beta _p}).C.U(B,{\beta _p})...U(C,{\gamma _1})\left| s \right\rangle \\
	= \frac{1}{{{2^n}}}{\left| {{f_1}(\vec \gamma ,\vec \beta )} \right|^2}P(\left| {00...0} \right\rangle ) + ... + \frac{1}{{{2^n}}}{\left| {{f_n}(\vec \gamma ,\vec \beta )} \right|^2}P(\left| {11...1} \right\rangle )
\end{array}
\end{equation}
As one can see, here, the eigenvalues of the objective function $C$ act as markers (eigenvalues of the solutions will be higher in numerical value), and by just demanding the maximum of the expectation value, the coefficients of the solutions ${\left| {{f_i}(\vec \gamma ,\vec \beta)} \right|^2}$ will be maximised for a given set of parameters ${\vec \gamma ,\vec \beta }$. Now, one can call the quantum computer to create the $\left| {\vec \gamma ,\vec \beta } \right\rangle $ and to obtain a string $z$.

The reason for the $p$-times iterations is that there is no promise that, after $\left| {{\gamma _1},{\beta _1}} \right\rangle  = U(B,{\beta _1})U(C,{\gamma _1})\left| s \right\rangle $, those coefficients related to the solutions in the above become maximised and the quantum state $\left| {{\gamma _1},{\beta _1}} \right\rangle $ becomes close enough to the solution $\left| {z'} \right\rangle $. Inevitably, we need to increase the degrees of freedom in the problem by $2p$ and hope that $p$ grows slower than $n$, thus becoming able to maximise the expectation value.    

The workflow of the QAOA started with determining the $2p$ angles $(\vec \gamma ,\vec \beta)$ that maximise the expectation value $F$; then, we constructed our quantum state $\left| {\vec \gamma ,\vec \beta } \right\rangle $ on the quantum computer and measured it on the computational basis to obtain the value $\left| z \right\rangle $. Finally, we calculated $C(z)$ on a classical computer. By repeating these steps, we obtained a distribution on $C(z)$, whose mean is $F(\vec \gamma ,\vec \beta)$.

One superior way for the optimisation of the $2p$ parameters ${\vec \gamma ,\vec \beta }$ is utilising the Gibbs objective function defined by $ - \log \left\langle {{e^{ - \eta C}}} \right\rangle $  \cite{7}.

\section{QAOA Implementation on MaxCut problems}

The MaxCut problem, one of the 21 NP-complete 
problems identified by Karp, has many applications in computer science and statistical physics. The MaxCut problem is in the APX 
class, telling us that, unless P = NP, no polynomial time approximation scheme is available for it, arbitrarily near to the optimal solution.

Herein, we first set the stage for the MaxCut problem, and then, by the explicit representation of its constraints as operators, we reviewed the construction of its related quantum gates in the QAOA algorithm. 

The MaxCut problem can be defined by $n$ vertices and $m$ edges. By introducing a two-valued function
that acts on the vertices, $k(i) \in \{  + 1, - 1\} $, a cut can be defined if $k(i)k(j) =  - 1$ for two vertices of an edge.

\begin{figure}
	\centering
\includegraphics[width=3 in]{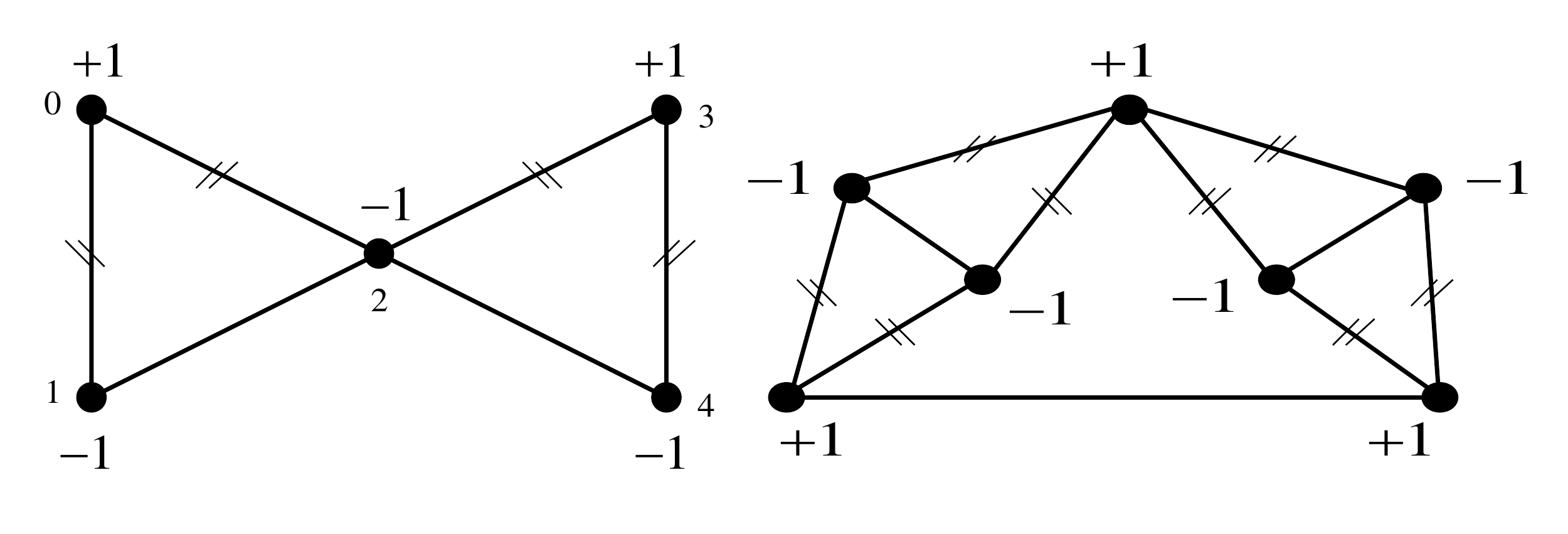}
	\caption{The MaxCut problem on the butterfly and Moser spindle graphs.}

\end{figure}

Thus, each vertex can be assigned by either  $+ 1 $ or  $ - 1$. The aim is to find a configuration of assignments, leading to a maximum number of cuts in the graph.

In order to solve such a problem on a quantum computer, we promote that the $n$ vertices be $n$ qubits. Thus, all different ${2^n}$ assignments for the vertices will correspond to the ${2^n}$ dimensional Hilbert space generated by $n$ qubits.

Now, one can easily define the objective function in the form of a constrained satisfaction problem (CSP).

\begin{equation}
\begin{array}{l}
C = \sum\limits_{ < ij > }^m {{C_{ < ij > }}} \\
{C_{ < ij > }} = \frac{1}{2}( + 1 - k(i)k(j))
\end{array}
\end{equation}

By promoting the objective function as an operator, we can get:

\begin{equation}
C\left| z \right\rangle  = \sum\limits_{ < ij > }^m {{C_{ < ij > }}} \left| z \right\rangle  = P(z)\left| z \right\rangle 
\end{equation}

where ${C_{ < ij > }} = \frac{1}{2}(+ I - \sigma _i^z \otimes \sigma _j^z)$.

Here, if $\left\langle {ij} \right\rangle $ is a cut, the related qubits will land onto different bases, for example, $\left| {{z_i}} \right\rangle  = \left| 0 \right\rangle $ and $\left| {{z_j}} \right\rangle  = \left| 1 \right\rangle $, or vice versa. Now, consider ${\sigma _i}^z \otimes {\sigma _j}^z\left| {01} \right\rangle  =  - \left| {01} \right\rangle $, then we can get ${C_{ < ij > }}\left| {{z_i}{z_j}} \right\rangle  = \left| {{z_i}{z_j}} \right\rangle $.

Now, we are ready to construct $U(C,\gamma)$ for the MaxCut so that:

\begin{equation}
U(C,\gamma ) = {e^{-i\gamma \sum\limits_{ < ij > } {{C_{ < ij > }}} }} = \prod\limits_{ < ij > } {{e^{ - i\frac{\gamma }{2}( + I - {\sigma _i}^z \otimes {\sigma _j}^z)}}} 
\end{equation}

Explicitly, on $\left| \psi  \right\rangle  = \left| {{y_1}} \right\rangle \left| {{y_2}} \right\rangle $.
\begin{equation}
{e^{ - i\frac{\gamma }{2}I \otimes I}}{e^{ + i\frac{\gamma }{2}{\sigma _1}^z \otimes {\sigma _2}^z}}\left| {{y_1}{y_2}} \right\rangle  = {\alpha _0}{\beta _0}\left| {00} \right\rangle  + {\alpha _0}{\beta _1}{e^{ - i\gamma }}\left| {01} \right\rangle  + {\alpha _1}{\beta _0}{e^{ - i\gamma }}\left| {10} \right\rangle  + {\alpha _1}{\beta _1}\left| {11} \right\rangle 
\end{equation}

The circuit for the above gate can be implemented easily with the help of one controlled-phase gate and two ordinary phase gates with the below arrangement:

\begin{equation}
\begin{array}{l}
CR( - 2\gamma )\left| {{y_1}{y_2}} \right\rangle  = {\alpha _0}{\beta _0}\left| {00} \right\rangle  + {\alpha _0}{\beta _1}\left| {01} \right\rangle  + {\alpha _1}{\beta _0}\left| {10} \right\rangle  + {\alpha _1}{\beta _1}{e^{ + i2\gamma }}\left| {11} \right\rangle \\
(R(\gamma ) \otimes R(\gamma ))CR( - 2\gamma )\left| {{y_1}{y_2}} \right\rangle  = {\alpha _0}{\beta _0}\left| {00} \right\rangle  + {\alpha _0}{\beta _1}{e^{ - i\gamma }}\left| {01} \right\rangle  + {\alpha _1}{\beta _0}{e^{ - i\gamma }}\left| {10} \right\rangle  + {\alpha _1}{\beta _1}\left| {11} \right\rangle 
\end{array}
\end{equation}

Now, the angle-dependent quantum state can be considered.

\begin{equation}
\left| {\vec \gamma ,\vec \beta } \right\rangle  = U(B,{\beta _p})U(C,{\gamma _p})...U(B,{\beta _1})U(C,{\gamma _1})\left| s \right\rangle 
\end{equation}

In order to determine the angles $\vec \gamma  = ({\gamma _1},...,{\gamma _p})$ and $\vec \beta  = ({\beta _1},...,{\beta _p})$, we consider the expectation value

\begin{equation}
F(\vec \gamma ,\vec \beta ) = \left\langle {\vec \gamma ,\vec \beta } \right|C\left| {\vec \gamma ,\vec \beta } \right\rangle  = \sum\limits_{ < ij > }^{} {\left\langle s \right|{U^\dag }(C,{\gamma _1})...{U^\dag }(B,{\beta _p}){C_{ < ij > }}U(B,{\beta _p})...U(C,{\gamma _1})\left| s \right\rangle } 
\end{equation}

Consider the operator associated with the edge $\left\langle {ij} \right\rangle $ for the case of $p=1$:
\begin{equation}
{U^\dag }(C,{\gamma _1}){U^\dag }(B,{\beta _1}){C_{ < ij > }}U(B,{\beta _1})U(C,{\gamma _1})
\end{equation}

The factors in ${U }(B,{\beta _1})$ which do not involve qubits i or j will commute with ${C_{ < ij > }}$ and we have

\begin{equation}
{U^\dag }(C,{\gamma _1}){e^{ + i{\beta _1}(\sigma _i^x + \sigma _j^x)}}{C_{ < ij > }}{e^{ - i{\beta _1}(\sigma _i^x + \sigma _j^x)}}U(C,{\gamma _1})
\end{equation}

Moreover, the factors in $U(C,{\gamma _1})$ that do not involve qubits $i$ or $j$ commute with ${e^{ - i{\beta _1}(\sigma _i^x + \sigma _j^x)}}$ and obviously with ${C_{ < ij > }}$, and then cancel out.

Thus, the operator in (23) 
only involves the edge $\left\langle {ij} \right\rangle $ and the edges adjacent to $\left\langle {ij} \right\rangle $, as well as the qubits on those edges.

Now, consider $p=2$, and we can obtain the operator:

\begin{equation}
{U^\dag }(C,{\gamma _1}){U^\dag }(B,{\beta _1}){U^\dag }(C,{\gamma _2}){U^\dag }(B,{\beta _2}){C_{ < ij > }}U(B,{\beta _2})U(C,{\gamma _2})U(B,{\beta _1})U(C,{\gamma _1})
\end{equation}

Here, the factors in $U(B,{\beta _2})$ that do not involve qubits $i$ and $j$ commute with ${C_{ < ij > }}$, and we can obtain:

\begin{equation}
{U^\dag }(C,{\gamma _1}){U^\dag }(B,{\beta _1}){U^\dag }(C,{\gamma _2}){e^{ + i{\beta _2}(\sigma _i^x + \sigma _j^x)}}{C_{ < ij > }}{e^{ - i{\beta _2}(\sigma _i^x + \sigma _j^x)}}U(C,{\gamma _2})U(B,{\beta _1})U(C,{\gamma _1})
\end{equation}

Now, the factors in $U(C,{\gamma _2})$ that will survive will be:

\begin{equation}
\prod\limits_{ < ik > , < jl > } {{e^{ - i\frac{{{\gamma _2}}}{2}( - \sigma _j^z\sigma _l^z + I)}}{e^{ - i\frac{{{\gamma _2}}}{2}( - \sigma _i^z\sigma _k^z + I)}}{e^{ - i\frac{{{\gamma _2}}}{2}( - \sigma _i^z\sigma _j^z + I)}}} 
\end{equation}

Obviously, the factors in $U(B,{\beta _1})$ that will survive will be ${e^{ - i{\beta _1}(\sigma _i^x + \sigma _j^x + \sigma _k^x + \sigma _l^x)}}$.

Finally, the factors in $U(C,{\gamma _1})$ that will survive will be:

\begin{equation}
\prod\limits_{ < ik > , < jl > , < km > , < \ln  > } {{e^{ - i\frac{{{\gamma _1}}}{2}( - \sigma _l^z\sigma _n^z + I)}}{e^{ - i\frac{{{\gamma _1}}}{2}( - \sigma _m^z\sigma _k^z + I)}}{e^{ - i\frac{{{\gamma _1}}}{2}( - \sigma _j^z\sigma _l^z + I)}}{e^{ - i\frac{{{\gamma _1}}}{2}( - \sigma _i^z\sigma _k^z + I)}}{e^{ - i\frac{{{\gamma _1}}}{2}( - \sigma _i^z\sigma _j^z + I)}}} 
\end{equation}

Thus, for $p=2$, the operator only involves edges that are, at most, two steps away from $\left\langle {ij} \right\rangle $.

Consequently, for any $p$, our operator only involves edges that are, at most, $p$ steps away from $\left\langle {ij} \right\rangle $ \cite{1}. 

Therefore, each term in Equation (22) depends only on the subgraph involving qubits $i$ and $j$ and those qubits at a distance less than or equal to $p$ steps away from $\left\langle {ij} \right\rangle $.

Now, let us define the restricted objective function ${C_S}$ that lives on a subgraph $S$ as below:
\begin{equation}
{C_S} = \sum\limits_{ < ij >  \in S}^{} {{C_{ < ij > }}} 
\end{equation}

and the associated operators:
\begin{equation}
\begin{array}{l}
U({C_S},\gamma ) = {e^{ - i\gamma {C_S}}}\\
U({B_S},\beta ) = {e^{ - i\beta {B_S}}}
\end{array}
\end{equation}

Where ${B_S} = \sum\limits_{i \in S}^{} {\sigma _i^x} $.

Each edge $\left\langle {ij} \right\rangle $, with its related subgraph, generates the below contribution to $F(\vec \gamma ,\vec \beta)$:
\begin{equation}
\left\langle {s,S(i,j)} \right|{U^\dag }({C_{S(i,j)}},{\gamma _1})...{U^\dag }({B_{S(i,j)}},{\beta _p}){C_{ < ij > }}U({B_{S(i,j)}},{\beta _p})...U({C_{S(i,j)}},{\gamma _1})\left| {s,S(i,j)} \right\rangle 
\end{equation}

It is easy to see that the contribution to the expectation value for different edges with isomorphic subgraphs are same. Thus, we can rewrite the expectation value as a sum over subgraph types.

Therefore, we can define:
\begin{equation}
\begin{array}{l}
{f_S}(\vec \gamma ,\vec \beta ) = \langle s,S(i,j)|{U^\dag }({C_{S(i,j)}},{\gamma _1})...{U^\dag }({B_{S(i,j)}},{\beta _p}){C_{ < ij > }}U({B_{S(i,j)}},{\beta _p})\\
...U({C_{S(i,j)}},{\gamma _1})\left| {s,S(i,j)} \right\rangle 
\end{array}
\end{equation}

And rewrite F as:
\begin{equation}
F(\vec \gamma ,\vec \beta ) = \sum\limits_S^{} {{m_S}} {f_S}(\vec \gamma ,\vec \beta )
\end{equation}

where ${{m_S}}$ is the number of occurrences of the subgraph S \cite{1}.

Regarding to the expectation value for MaxCut problem, we have:
\begin{equation}
\frac{m}{2} + \frac{{ - 1}}{2}\sum\limits_{ < ij > }^m {Tr[{M_d}.\prod\limits_{ < ij > }^m {{e^{ + i\frac{\gamma }{2}( + I - \sigma _i^z\sigma _j^z)}}} .{e^{ + i\beta (\sigma _i^x + \sigma _j^x)}}.(\sigma _i^z\sigma _j^z).{e^{ - i\beta (\sigma _i^x + \sigma _j^x)}}.\prod\limits_{ < ij > }^m {{e^{ - i\frac{\gamma }{2}( + I - \sigma _i^z\sigma _j^z)}}} ]} 
\end{equation}

Where ${M_d}$ is the initial density matrix.

By considering $U(B,\beta ) = \prod\limits_i^n {({\mathop{\rm Cos}\nolimits} (} \beta )I - i{\mathop{\rm Sin}\nolimits} (\beta )\sigma _i^x)$, we can simplify the above trace as:

\begin{equation}
\begin{array}{l}
	{\mathop{\rm Cos}\nolimits} (2\beta ){\mathop{\rm Sin}\nolimits} (2\beta )Tr[{M_d}.\prod\limits_{ < ij > }^m {{e^{ + i\frac{\gamma }{2}( + I - \sigma _i^z\sigma _j^z)}}} .(\sigma _i^z\sigma _j^y).\prod\limits_{ < ij > }^m {{e^{ - i\frac{\gamma }{2}( + I - \sigma _i^z\sigma _j^z)}}} ]\\
	+ {\mathop{\rm Cos}\nolimits} (2\beta ){\mathop{\rm Sin}\nolimits} (2\beta )Tr[{M_d}.\prod\limits_{ < ij > }^m {{e^{ + i\frac{\gamma }{2}( + I - \sigma _i^z\sigma _j^z)}}} .(\sigma _i^y\sigma _j^z).\prod\limits_{ < ij > }^m {{e^{ - i\frac{\gamma }{2}( + I - \sigma _i^z\sigma _j^z)}}} ]\\
	+ {\mathop{\rm Sin}\nolimits} {(2\beta )^2}Tr[{M_d}.\prod\limits_{ < ij > }^m {{e^{ + i\frac{\gamma }{2}( + I - \sigma _i^z\sigma _j^z)}}} .(\sigma _i^y\sigma _j^y).\prod\limits_{ < ij > }^m {{e^{ - i\frac{\gamma }{2}( + I - \sigma _i^z\sigma _j^z)}}} ]
\end{array}
\end{equation}

Moreover, one can derive an analytical expression for the expectation value by further decomposing and simplifying these traces.

\section{QAOA Implementation on QUBOs and Ising Type Hamiltonians}
One of the main applications of quantum computers is solving NP-hard combinatorial optimisation problems efficiently. With classical computation paradigms, in most of the cases, all we have are heuristic algorithms with which to tackle these computationally intensive problems. Therefore, it worth seeking quantum algorithms with the ability to find higher-quality solutions for such high-valued problems. A wide range of real-world optimisation problems can be recast into higher order binary optimisation (HOBO) and quadratic unconstrained binary optimisation problems (QUBOs) \cite{8,9}.
In the case of HOBO, the natural prescription is to transform a higher-order problem into a quadratic one (${\rm{quadratisation}}$)\cite{10,11}. Moreover, for solving very large versions of this problem, it has been shown that decomposing large QUBO instances into mini-QUBOs and merging the sub-solutions can provide an effective solution \cite{12,13,14}.
QUBOs can be connected to Ising Hamiltonian via linear transformations. Here, we apply the QAOA to a Ising Hamiltonian. This will pave the way for solving QUBOs on gate-based quantum computers.

The objective function in QUBO formulation can be written as:

\begin{equation}
{H_{QUBO}} = \sum\limits_{i < j} {{J_{ij}}} {x_i}{x_j} + \sum\limits_i {{J_i}} {x_i} = \sum\limits_{i \le j}^{} {{x_i}} {J_{ij}}{x_j} = \left\langle {\vec x,J\vec x} \right\rangle  
\end{equation}
The variables ${x_i}$ live in the Boolean space $\{ 0,1\} $.
Usually, the ${{J_{ij}}}$ matrix is written in symmetric or in upper triangle form.

In the Ising model, the objective function can be written as:
\begin{equation}
{H_{Ising}} = \sum\limits_{i < j} {{J_{ij}}} {s_i}{s_j} + \sum\limits_i {{J_i}} {s_i}
\end{equation}
Here the variables ${s_i}$ live in the Ising space $\{ -1,1\} $.

Translation between Ising and QUBO representations can be done with simple mapping, $s = 2x - 1$, where 1 is the vector all of whose components are 1.

With the above linear transformation, one can establish a simple relationship between the two objective functions of QUBO and Ising.
\begin{equation}
\left\langle {\vec x,J\vec x} \right\rangle  = \frac{1}{4}(\left\langle {\vec 1,J\vec 1} \right\rangle  + \left\langle {\vec s,J\vec s} \right\rangle  + 2\left\langle {\vec 1,J\vec s} \right\rangle )
\end{equation}

Now, by promoting binary variables ${s_i}$ to Pauli matrices ${\sigma _i}^z$, we can achieve a quantum Ising Hamiltonian as:
\begin{equation}
H = \sum\limits_{i < j} {{J_{ij}}} {\sigma _i}^z{\sigma _j}^z + \sum\limits_{i = 1}^n {{J_i}} {\sigma _i}^z
\end{equation}

where it is understood that:
\begin{equation}
{\sigma _i}^z{\sigma _j}^z = {{\rm I}_1} \otimes ... \otimes {{\rm I}_{i - 1}} \otimes {\sigma _i}^z \otimes {{\rm I}_{i + 1}} \otimes ... \otimes {{\rm I}_{j - 1}} \otimes {\sigma _j}^z \otimes {{\rm I}_{j + 1}} \otimes ... \otimes {{\rm I}_n}
\end{equation}

We can construct $U(H,\gamma)$, so that:

\begin{equation}
U(H,\gamma ) = {e^{ - i\gamma H}} = \prod\limits_{i < j} {{e^{ - i\gamma {J_{ij}}{\sigma _i}^z{\sigma _j}^z}}} \prod\limits_i {{e^{ - i\gamma {J_i}{\sigma _i}^z}}} 
\end{equation}

The quantum gate $U(B,\beta)$ should be the same as (11). Now, we can construct the variational ansatz as:

\begin{equation}
\left| {\vec \gamma ,\vec \beta } \right\rangle  = U(B,{\beta _p})U(H,{\gamma _p})...U(B,{\beta _1})U(H,{\gamma _1}){\left|  +  \right\rangle ^{ \otimes n}}
\end{equation}

Where ${\left|  +  \right\rangle ^{ \otimes n}}$ is the uniform superposition, can be constructed via applying the Hadamard gates ${H^1} \otimes ... \otimes {H^n}$ to ${\left| 0 \right\rangle ^{ \otimes n}}$ and note $\vec \gamma  = ({\gamma _1},...,{\gamma _p})$ and $\vec \beta  = ({\beta _1},...,{\beta _p})$.

In order to run the quantum algorithm, ${\vec \gamma }$ and ${\vec \beta }$ need to be determined by minimisation (or maximisation) of the expectation value of the Hamiltonian as below:
\[H(\vec \gamma ,\vec \beta ) = \left\langle {\vec \gamma ,\vec \beta } \right|H\left| {\vec \gamma ,\vec \beta } \right\rangle \]

For the simplest case, $p = 1$, we arrive at:
\begin{equation}
\left\langle H \right\rangle  = \left\langle s \right|{U^\dag }(H,{\gamma _1}){U^\dag }(B,{\beta _1})(\sum\limits_{i < j} {{J_{ij}}} {\sigma _i}^z{\sigma _j}^z + \sum\limits_{i = 1}^n {{J_i}} {\sigma _i}^z)U(B,{\beta _1})U(H,{\gamma _1})\left| s \right\rangle 
\end{equation}

For $p = 1$ and for a triangle-free graph, one can check that the below factors will survive:
\begin{equation}
\begin{array}{l}
\left\langle H \right\rangle  = \left\langle {{H_1}} \right\rangle  + \left\langle {{H_2}} \right\rangle \\
= \sum\limits_i^{} {^n\left\langle  +  \right|} \prod\limits_{(i,j) \in E} {{e^{ + i\gamma {J_{ij}}{\sigma _i}^z{\sigma _j}^z}}} {e^{ + i\gamma {J_i}{\sigma _i}^z}}{e^{ + i\beta {\sigma _i}^x}}({J_i}{\sigma _i}^z){e^{ - i\beta {\sigma _i}^x}}{e^{ - i\gamma {J_i}{\sigma _i}^z}}\prod\limits_{(i,j) \in E} {{e^{ - i\gamma {J_{ij}}{\sigma _i}^z{\sigma _j}^z}}} {\left|  +  \right\rangle ^n}\\
+ \sum\limits_{(i,j)}^{} {^n\left\langle  +  \right|} (\prod\limits_{(j,l) \in E} {{e^{ + i\gamma {J_{jl}}{\sigma _j}^z{\sigma _l}^z}})} (\prod\limits_{(i,k) \in E} {{e^{ + i\gamma {J_{ik}}{\sigma _i}^z{\sigma _k}^z}}} )({e^{ + i\gamma {J_{ij}}{\sigma _i}^z{\sigma _j}^z}})(\prod\limits_{i,j} {{e^{ + i\gamma {J_i}{\sigma _i}^z}})(} (\prod\limits_{i,j} {{e^{ + i\beta {\sigma _i}^x}})} ({J_{ij}}{\sigma _i}^z{\sigma _j}^z)\\
(\prod\limits_{i,j} {{e^{ - i\beta {\sigma _i}^x}})} (\prod\limits_{i,j} {{e^{ - i\gamma {J_i}{\sigma _i}^z}})(} {e^{ - i\gamma {J_{ij}}{\sigma _i}^z{\sigma _j}^z}})(\prod\limits_{(i,k) \in E} {{e^{ - i\gamma {J_{ik}}{\sigma _i}^z{\sigma _k}^z}}} )(\prod\limits_{(j,l) \in E} {{e^{ - i\gamma {J_{jl}}{\sigma _j}^z{\sigma _l}^z}})} {\left|  +  \right\rangle ^n}
\end{array}
\end{equation}

The expectation value can be achieved analytically as:

\begin{equation}
\begin{array}{l}
\left\langle H \right\rangle  = \sum\limits_i^{} {{J_i}} \sin (2\beta )\sin (2\gamma {J_i})\prod\limits_{(i,j) \in E} {\cos (2\gamma {J_{ij}})} \\
+\sum\limits_{(i,j)}^{} {\frac{1}{2}{{\rm{J}}_{ij}}\sin (\beta )\cos (\beta )(2\cos (2\beta )\sin (2\gamma {{\rm{J}}_{ij}})(} 2\cos (2\gamma {{\rm{J}}_i})\prod\limits_{(i,k)} {\cos (2\gamma {{\rm{J}}_{ik}})} \\
+ 2\cos (2\gamma {{\rm{J}}_{\rm{j}}})\prod\limits_{(j,l)} {\cos (2\gamma {{\rm{J}}_{jl}})} )\\
+ 4\sin (2\beta )\sin (2\gamma {{\rm{J}}_i})\sin (2\gamma {{\rm{J}}_j})\prod\limits_{(i,k)} {\cos (2\gamma {J_{ik}})} \prod\limits_{(j,l)} {\cos (2\gamma {{\rm{J}}_{jl}})} )
\end{array}
\end{equation}

\section{Quantum Algorithm Implementations}
Here, we review the implementation of the QAOA on some explicit examples (MaxCut on butterfly and Moser spindle graphs). In order to find the optimal angles $\vec \gamma$ and $\vec \beta$, especially for cases of $p>1$, different prescriptions have been offered, such as gradient descent and basin-hopping \cite{15,16}. Such treatments on the QAOA need to be automated to have ready-to-run QAOA jobs at peak performance.

\begin{figure}
	\centering
	\includegraphics[width=5in]{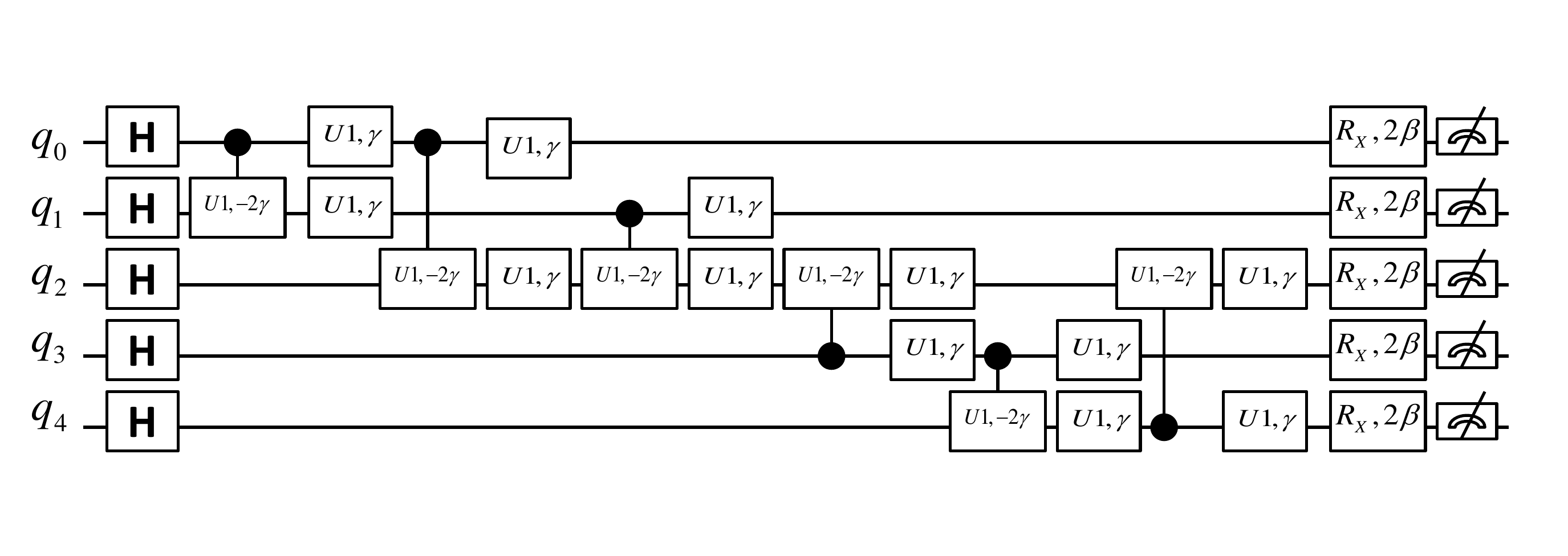}
	\caption{QAOA circuit implementation for MaxCut problem on butterfly graph}
	
\end{figure}

Returning to the MaxCut problem on a butterfly graph, for $p = 1$, the expectation value is:

\begin{equation}
F(\gamma ,\beta ) = \left\langle s \right|{U^\dag }(C,\gamma ){U^\dag }(B,\beta ).C.U(B,\beta )U(C,\gamma )\left| s \right\rangle 
\end{equation}

Based on the simplification procedure discussed in the previous section, we can write $F(\gamma ,\beta) = 2{f_A} + 4{f_B}$, where ${f_A}$ and ${f_B}$ are the expectation values for one edge in the sub-graphs $\{ (0,1),(3,4)\} $ and $\{ (0,2),(1,2),(3,2),(4,2)\} $, respectively.

\begin{equation}
\begin{array}{l}
{f_A} = \frac{1}{2}(1 - \left\langle {{ + ^3}} \right|U_{12}^\dag (\gamma )U_{02}^\dag (\gamma )U_{01}^\dag (\gamma )U_1^\dag (\beta )U_0^\dag (\beta )({\sigma _0}^z \otimes {\sigma _1}^z)\\
{U_0}(\beta ){U_1}(\beta ){U_{01}}(\gamma ){U_{02}}(\gamma ){U_{12}}(\gamma )\left| {{ + ^3}} \right\rangle )
\end{array}
\end{equation}

\begin{equation}
{f_A} = \frac{1}{4}(\sin (2\beta )\sin (\gamma )(\cos (2\beta  - \gamma ) + 3\cos (2\beta  + \gamma )) + 2)
\end{equation}

\begin{equation}
\begin{array}{l}
{f_B} = \frac{1}{2}(1 - \left\langle {{ + ^5}} \right|U_{24}^\dag (\gamma )U_{23}^\dag (\gamma )U_{12}^\dag (\gamma )U_{02}^\dag (\gamma )U_{01}^\dag (\gamma )U_2^\dag (\beta )U_0^\dag (\beta )({\sigma _0}^z \otimes {\sigma _2}^z)\\
{U_0}(\beta ){U_2}(\beta ){U_{01}}(\gamma ){U_{02}}(\gamma ){U_{12}}(\gamma ){U_{23}}(\gamma ){U_{24}}(\gamma )\left| {{ + ^5}} \right\rangle )
\end{array}
\end{equation}

\begin{equation}
{f_B} = \frac{1}{8}(\sin (2\beta )\sin (2\gamma )(\cos (2(\beta  + \gamma )) + 3\cos (2\beta )) + 4)
\end{equation}

The six gates of form $U({C_{ij}},\gamma)$ can be written as below using phase shift gates $R(\gamma)$ and controlled phase shift gates $CR(- 2\gamma)$.

\begin{equation}
U({C_{ij}},\gamma ) = {e^{ - i\frac{\gamma }{2}( - \sigma _i^z \otimes \sigma _j^z + I)}} = ({R_i}(\gamma ) \otimes {R_j}(\gamma )).{C_{ij}}R( - 2\gamma )
\end{equation}

\textbf{Classical Preprocessing to Determine $\gamma$ and $\beta$}

\begin{verbatim}
#In Python, set up your Wolfram language session:
from wolframclient.language import wlexpr
from wolframclient.evaluation import WolframLanguageSession
session = WolframLanguageSession()
#Evaluate any Wolfram Language code from Python:
Angles=session.evaluate(wlexpr('NArgMax[1/ 8 (21 + 
3 Cos[4 \[Beta]] + 4 Cos[4 \[Beta] - 2 \[Gamma]] +  
2 Cos[2 \[Gamma]] + Cos[4 \[Gamma]] -  Cos[4 (\[Beta] 
+ \[Gamma])] -  6 Cos[2 (2 \[Beta] + \[Gamma])]), 
{\[Gamma], \[Beta]} \[Element]  
DiscretizeRegion[Disk[{0, 0}, {2 Pi, Pi}, {0, Pi/2}]]]'))

Gamma=Angles[0]
Beta=Angles[1]
\end{verbatim}

\begin{verbatim}
# importing Qiskit
from qiskit import QuantumRegister, ClassicalRegister, QuantumCircuit
from qiskit import IBMQ, execute
from qiskit.tools.monitor import job_monitor
from qiskit.visualization import plot_histogram
\end{verbatim}

\begin{figure}
	\centering
	\includegraphics[width=5in]{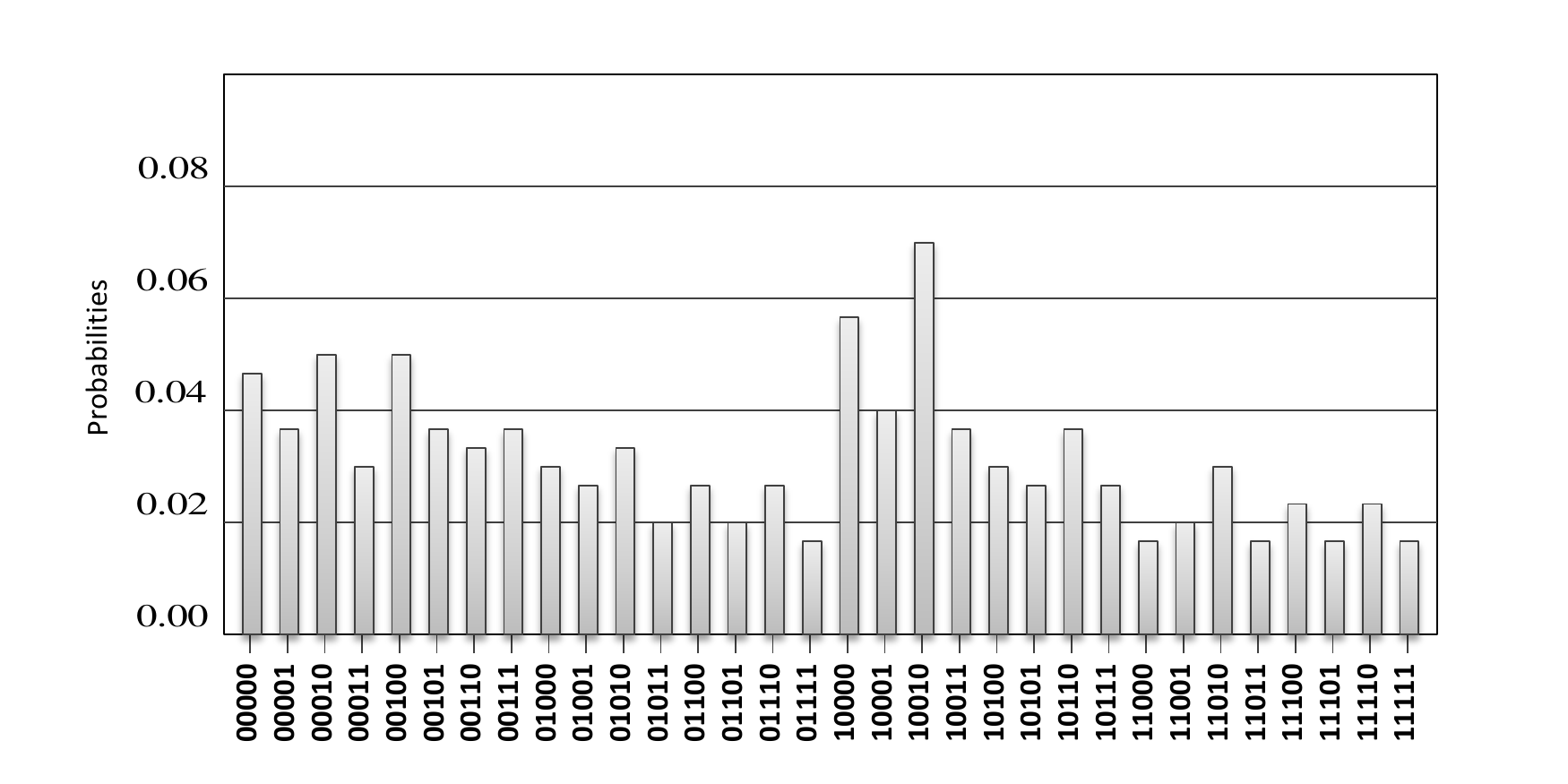}
	\caption{Running the QAOA on ibmq-16-melbourne device} 
	
\end{figure}

\begin{verbatim}
# prepare the butterfly graph [17]
import numpy as np
m = 5
Nodes = np.arange(0,m,1)
Edges =[(0,1),(0,2),(1,2),(3,2),(3,4),(4,2)]
# prepare the quantum and classical resisters
QAlgo = QuantumCircuit(len(Nodes), len(Nodes))
# apply the Hadamard gates to all qubits to have uniform superposition state
QAlgo.h(range(len(Nodes)))
QAlgo.barrier()
# to implement U(C,gamma), apply below equivalent circuit [17]
# note here u1 is the phase shift gate
for edge in Edges:
    i = edge[0]
    j = edge[1]
    QAlgo.cu1(-2*Gamma, i, j)
    QAlgo.u1(Gamma, i)
    QAlgo.u1(Gamma, j)

# to implement U(B,beta), apply R_x(2*beta) to all qubits
QAlgo.barrier()
QAlgo.rx(2*Beta, range(len(Nodes)))
# Measure the result in the computational basis
QAlgo.barrier()
QAlgo.measure(range(len(Nodes)),range(len(Nodes)))
\end{verbatim}

\begin{verbatim}
# Use the ibmq_16_melbourne device
from qiskit.tools.monitor import job_monitor
provider = IBMQ.load_account()
backend = provider.get_backend('ibmq_16_melbourne')
shots = 2048
job = execute(QAlgo, backend=backend, shots=shots)
job_monitor(job)
results = job.result() 
plot_histogram(results.get_counts(),bar_labels = False,figsize = (12,10))
\end{verbatim}

For the Moser spindle graph, one can directly compute the single-layer expectation value and the related angles from the below in Wolfram language:

\begin{verbatim}

MaxCut[x_, y_] := {NE = {Range[x], y}, nodes = NE[[1]], 
edges = NE[[2]], cons = Range[Length[edges]],
Sig[i_, j_] := ReplacePart[Table[IdentityMatrix[2], 
{k, Length[NE[[1]]]}], {i -> PauliMatrix[3], j -> PauliMatrix[3]}],
Do[cons[[i]] = Sig[edges[[i]][[1]], edges[[i]][[2]]], 
{i, Length[edges]}],consyy = Range[Length[edges]],Sigyy[i_, j_] := 
ReplacePart[Table[IdentityMatrix[2], {k, Length[NE[[1]]]}], 
{i -> PauliMatrix[2], j -> PauliMatrix[2]}],
Do[consyy[[i]] = Sigyy[edges[[i]][[1]], edges[[i]][[2]]], 
{i, Length[edges]}],consizyj = Range[Length[edges]],Sigizyj[i_, j_] := 
ReplacePart[Table[IdentityMatrix[2], {k, Length[NE[[1]]]}], {i -> 
PauliMatrix[3], j -> PauliMatrix[2]}],
Do[consizyj[[i]] = Sigizyj[edges[[i]][[1]], edges[[i]][[2]]], 
{i, Length[edges]}],consiyzj = Range[Length[edges]],
Sigiyzj[i_, j_] := ReplacePart[Table[IdentityMatrix[2], 
{k, Length[NE[[1]]]}], {i -> PauliMatrix[2], j -> PauliMatrix[3]}],
Do[consiyzj[[i]] = Sigiyzj[edges[[i]][[1]], edges[[i]][[2]]], 
{i, Length[edges]}],Conjugate[\[Gamma]] ^:= \[Gamma],
Mapi[i_] := Function[k, Exp[-I \[Gamma] (1/2) k]] /@ 
Diagonal[(-KroneckerProduct @@ cons[[i]] + 
IdentityMatrix[2^Length[nodes]])],
UCgamma = DiagonalMatrix[Product[Mapi[i], {i, Length[edges]}]],
DaggerUCgamma = ConjugateTranspose[UCgamma],
reducedCijyy[i_] := KroneckerProduct @@ consyy[[i]],
reducedCizyj[i_] := KroneckerProduct @@ consizyj[[i]],
reducedCiyzj[i_] := KroneckerProduct @@ consiyzj[[i]],
Do[reducedCijyy[i]; reducedCizyj[i]; reducedCiyzj[i], 
{i, Length[edges]}],Conjugate[\[Beta]] ^:= \[Beta],
DenMat = Table[1/2^Length[nodes], {i, 1, 2^Length[nodes]}, 
{j, 1, 2^Length[nodes]}],
TRACE[i_] := Cos[2 \[Beta]] Sin[2 \[Beta]] Diagonal[
DenMat.DaggerUCgamma.reducedCizyj[i]].Diagonal[UCgamma] + 
Cos[2 \[Beta]] Sin[2 \[Beta]] Diagonal[DenMat.DaggerUCgamma.
reducedCiyzj[i]].Diagonal[UCgamma] + 
Sin[2 \[Beta]]^2 Diagonal[DenMat.DaggerUCgamma.
reducedCijyy[i]].Diagonal[UCgamma],
ParallelDo[TRACE[k], {k, Length[NE[[2]]]}], 
Expectval = FullSimplify[(1/2 Length[NE[[2]]] - 
1/2 ParallelSum[TRACE[i], {i, Length[NE[[2]]]}])],
Print["The Expectation Value: ", Expectval], 
Angles = NArgMax[Expectval, {\[Gamma], \[Beta]} \[Element] 
Disk[{0, 0}, {2 Pi, Pi}]]}
	
\end{verbatim}

\begin{verbatim}

MaxCut[7,{{1,2},{1,3},{2,3},{4,3},{2,4},{4,5},{6,5},{5,7},{6,7},{1,6},{1,7}}];
\end{verbatim}

In this example, the single layer expectation value is:
\begin{equation}
\frac{1}{{16}}\left( {88 + 8\cos {{\left[ \gamma  \right]}^2}\left( {9 + 2\cos \left[ \gamma  \right]} \right)\sin \left[ {4\beta } \right]\sin \left[ \gamma  \right] - 8\left( {2 + \cos \left[ \gamma  \right]} \right)\sin {{\left[ {2\beta } \right]}^2}\sin {{\left[ {2\gamma } \right]}^2}} \right)
\end{equation}

\section*{Acknowledgement}

The author would like to thank Nike Dattani and Edward Farhi for their helpful comments.

\end{document}